\documentclass[rnote]{aa}

\usepackage{graphicx,amsmath}
\usepackage{natbib}
\bibpunct{(}{)}{;}{a}{}{,}

\begin{document}

\title{Boundary conditions for NLTE polarized radiative transfer with incident radiation}
 \author{   M.\ Faurobert\inst{1},  I.  \ Mili\'{c}\inst{1,2}, O. \ Atanackovi\'{c}\inst{3}}
  \offprints{marianne.faurobert@unice.fr}
\institute{UMR 7293 Lagrange, Universit\'{e} de Nice Sophia Antipolis,
    CNRS, Observatoire de la C\^{o}te d'Azur, Campus Valrose, F-06108 Nice, France\\
\and Astronomical observatory Belgrade, Volgina 7, 11060 Belgrade, Serbia \\    
\and Department of Astronomy, Faculty of Mathematics, University of Belgrade, Studentski Trg 16, 11000 Belgrade, Serbia\\
    \email{marianne.faurobert@unice.fr}
    }
\date{Received/Accepted}
\titlerunning{Boundary conditions for polarized radiative transfer }
 \authorrunning{Faurobert {\it et al}}
\abstract
{Polarized NLTE radiative transfer in the presence of scattering in spectral lines and/or  in continua may be cast in a so-called reduced form for six reduced components of the radiation field. In this formalism the six components of the reduced source function are angle-independent quantities. It thus reduces drastically the storage requirement of numerical codes and it is  very well suited for solving polarized NLTE radiative transfer problems in 3D media.}
{This approach encounters a fundamental problem when the medium is illuminated by a polarized incident radiation, because there is a priori no way of relating the known (and measurable) Stokes parameters of the incident radiation to boundary conditions for the reduced equations. The origin of this problem is that there is no unique way of deriving the radiation reduced components from its Stokes parameters (only the inverse operation is clearly defined). The method proposed here aims at enabling to work with arbitrary incident
radiation field (polarized or unpolarized).}
{In previous works an  ad-hoc treatment of the boundary conditions, applying to cases where the incident radiation is unpolarized, has been used.
In this note we show that it is possible to account for the incident radiation in a rigorous way, without any assumption on its properties,  by expressing the radiation field as the sum of a directly 
transmitted radiation and of a so-called diffuse radiation.}
{The diffuse radiation field obeys a transfer equation with no incident radiation that may be solved in the reduced form. The first scattering of the incident radiation introduces  primary creation terms in the six components of the reduced source function. Once the reduced polarized transfer problem is solved for the diffuse radiation field, its Stokes parameters can be computed. The full radiation field is then obtained by adding the directly transmitted radiation field  computed in the Stokes formalism.}
{ In the case of an unpolarized incident radiation,  the diffuse field approach allows us to validate the previously introduced ad-hoc expressions.  The diffuse field approach however leads to more accurate 
computation of the source terms in the case where the incident radiation is anisotropic.   It is the only possible approach when the incident radiation field is polarized.
We perform numerical computations of  such cases, showing that the emergent line-polarization may be significantly affected by the polarization of the incident radiation.}

\keywords{}

\maketitle

\section{Introduction}
Solving NLTE polarized radiative transfer problems in multi-dimensional media is a challenge, mostly from a numerical point of view. It requires, in principle,
much more memory and computing time than the unpolarized problem, because the source function for polarized radiation depends
both on the spatial coordinates and on the two angles $(\theta, \varphi)$ which define the propagation direction of the radiation. Furthermore, a polarized radiation
is described by 4 measurable quantities, the so-called Stokes parameters (instead of one, in the unpolarized, scalar case) which are coupled by the physics of  
the absorption and emission processes. 

In this research note we limit ourselves to problems where the polarization arises from scattering processes and where the absorption of radiation 
does not introduce any coupling between the Stokes parameters. This is the case of line scattering problems when the Zeeman sub-levels of the line lower-level 
are evenly populated, and for Rayleigh or Thomson scattering problems in continua. We consider only linearly polarized radiation, as scattering processes
do not couple linear and  circular polarization. The Hanle effect due to the presence of weak magnetic fields is accounted for in the line scattering case.

The common point of these problems, from a mathematical point of view, is that the scattering  can be described by a so-called scattering matrix, which depends both on the directions and frequencies of the in-coming and out-going photons. It also depends on the magnetic field strength and direction, in the presence of Hanle effect.
A significant step toward the efficient solution of scattering problems in 3D media has been achieved thanks to the work of \citet{ anusha2011a, anusha2011c} and the subsequent
 series of papers by the same authors. Their work is based on the previous paper of \citet{frisch2007} who showed that the scattering phase matrix 
can be written as the sum of terms where the in-going and out-going directions are factorized. This factorization applies when the frequency redistribution  due to scattering
is independent of the directions of the photons. This is the case when the line is formed with complete frequency redistribution, whereas in the presence of 
partial frequency redistribution it relies on the assumption that the angle-averaged  frequency redistribution function may be used instead of the full angle-dependent one.
In this note,  we shall work in the framework of this assumption, but it could be relaxed following the same approach as in \citet{AnushaV}. 

The NLTE polarized radiative transfer equation for the three coupled Stokes parameters $I$, $Q$, and $U$ requires the computation of three 
source terms  depending on the position, frequency, and on two angles defining the radiation propagation direction.
We shall see in the following that the reduced
formalism allows us to deal with six reduced source terms depending on the position and frequency only. In multi-dimensional media
the radiation field may be locally highly anisotropic, and the anisotropy is an important source of polarization, fine angular grids are thus required for accurate 
numerical solutions of the polarized radiative transfer equation. The reduced formalism then  drastically reduces the memory and computing time requirements
of numerical solutions.

\section{Reduced formalism for polarized radiative transfer}
Let us summarize the main steps of the derivation leading to the reduced form of the polarized radiative transfer equation.
\subsection{Reduced radiative transfer equation}
We assume that  scattering phase matrix  may be written as a sum of factorized terms where the frequency and angular variables are separated, 
\begin{eqnarray}
{\hat R}(x,\Omega,x',\Omega',{\bf B})&=&[r_{II}(x,x'){\hat P}_{II}(\Omega,\Omega',{\bf B})\cr
& + & r_{III}(x,x'){\hat P}_{III}(\Omega,\Omega',{\bf B})],
\label{eq0}
\end{eqnarray}
where $r_{II}(x,x')$ and $r_{III}(x,x')$ are the standard angle-averaged redistribution functions corresponding to coherent scattering and 
to complete frequency redistribution in the atomic rest frame, respectively. The matrices ${\hat P}_{II}$ and ${\hat P}_{III}$ are given in the Appendix C of 
\citet{  anusha2011c}, according to the results established in  \citet{bommier97a, bommier97b}. They are related to the
Rayleigh and  Hanle phase matrices, which may be further factorized in terms of the irreducible spherical tensors ${\cal T}^ K_Q (i, \Omega) $ first introduced
  in \citet{landi84} \citep[see also][]{landibook}. 
  
  The Hanle phase matrix is given by
\begin{equation}
 P^H_{i,j}(\Omega,\Omega',{\bf B})=\sum_{K,Q}{\cal T}_Q^K(i,\Omega)\sum_{Q'}{\cal M}_{QQ'}^K({\bf B})(-1)^{Q'}{\cal T}_{-Q'}^K(j,\Omega'),
\label{eq0bis}
\end{equation}
where $\Omega$ and  $\Omega'$ denote the propagation directions of the out-going and in-going photons, respectively,  and $x$ and $x'$ are their frequencies 
measured with respect to the line central frequency in Doppler width unit, ${\bf B}$ is the magnetic field vector. The index $i= 0, 1, 2, 3$ refers to the four Stokes parameters.
The matrix ${\cal M}_{Q,Q'}^K(B)$ is given in \citet{landibook} (chap. 5).

In the absence of a magnetic field, the Rayleigh phase matrix  reduces to
\begin{equation}
 P^R_{i,j}(\Omega,\Omega')=\sum_{K,Q}{\cal T}_Q^K(i,\Omega)(-1)^{Q}{\cal T}_{-Q}^K(j,\Omega')
\label{eq0ter}
\end{equation}
The irreducible spherical tensors are well suited for handling the angular dependence of Rayleigh-type scattering phase matrices, with
$K$= 0,1,2 and $-K\le Q\le K$. In such problems, the circular and linear polarization decouple.

Using these factorizations \citet{ anusha2011a, anusha2011c} showed that the polarized radiative transfer equation  along the ray
for the Stokes vector ${\bf I} = (I,Q,U)^T$, namely, 
\begin{equation} 
  \frac{d {\bf I}( {\bf r},\Omega, x)}{ds}= -k_{tot}( {\bf r}, x) [ {\bf I}( {\bf r},\Omega, x) - {\bf S}( {\bf r},\Omega, x)],
 \label{eq1}  
\end{equation}
 may be cast into a set of six formally independent transfer equations for the 6-component  reduced Stokes vector 
 ${\cal I}= (I_0^0,I_{0}^2, I_{-1}^2,I_{1}^2,I_{-2}^2,I_{2}^2)^T$,
 where the corresponding 6-component
  reduced source vector  ${\cal S}({\bf r}, x)$ is angle-independent, i.e.
 \begin{equation} 
  \frac{d {\cal I}( {\bf r},\Omega, x)}{ ds}= - k_{tot}( {\bf r}, x) [ {\cal I}( {\bf r},\Omega, x) - {\cal S}( {\bf r}, x)],
 \label{eq2}  
\end{equation}
 
As usual $k_{tot}( {\bf r}, x)=k_l ( {\bf r})\phi(x) + k_c( {\bf r})$, where $k_l$ and $k_c$ denote respectively the line-integrated and 
the {\bf continuum} absorption coefficients, $\phi(x)$ is the normalized line absorption profile.
 The coordinate $s$ is the path length along the ray. 
 
 The reason for this remarkable property is  that the source vector in Eq. (\ref{eq1}) may be written in the irreducible spherical tensors basis as, 
 \begin{equation}
 S_i({\bf r}, \Omega, x)= \sum_{QK }{\cal T}^ K_Q (i, \Omega)S_Q^K({\bf r},x).
  \label{eq3}
 \end{equation}
 where the  irreducible components $S_Q^K({\bf r},x)$ of the total source vector  are independent of the propagation direction. 
 This is a direct consequence of the factorization of the scattering phase matrix. 
 
 We recall here that, as already noticed in \citet{frisch2007}, the reduced intensity components defined
  as the solutions of the reduced radiative transfer equations, are different from the tensors ${\cal I}_Q^K$ defined in \citet{landibook} 
 whose angle-averages appear in the statistical equilibrium equations for the element of the atomic density matrix.
 
 The  $S_Q^K$ components of the total source function are simply related to the reduced
 components of the line and continuum source functions $S_{l,Q}^K$ and $S_{c,Q}^K$, by 
  \begin{equation}
 S_Q^K({\bf r},x)=  \frac{k_l ( {\bf r})\phi(x)}{ k_{tot}}S_{l,Q}^K( {\bf r}, x)+{k_c( {\bf r})\over k_{tot}}S_{c,Q}^K( {\bf r}, x).
 \label{eq4}
 \end{equation}

 The reduced line source function is given in Eqs. (18-19) of \citet{anusha2011c} in terms of the reduced 6-component radiation vector ${\cal I}$, so we may write it as
  \begin{eqnarray}
 {\cal S}_l({\bf r},x) =   {\cal G}_l({\bf r}) 
  +{1\over\phi(x)}\int_{-\infty}^\infty & dx' & \int{d\Omega' \over 4\pi} {\hat R}(x,x', {\bf B})\cr
 &  &{\hat \Psi}(\Omega'){\cal I}( {\bf r},\Omega', x') ,
  \label{eq5}
 \end{eqnarray}
 where the 6x6 matrix $ {\hat \Psi}$ is given by
 \begin{equation}
  {\hat \Psi}_{Q,Q'}^{K,K'}(\Omega)=\sum _{j=0}^3(-1)^Q{\cal T}^ K_{-Q} (j, \Omega){\cal T}^ {K'}_{Q'} (j, \Omega). 
  \label{eq6}
  \end{equation}
In Eq.(\ref{eq5}) the first term of the right-hand side denotes the reduced components of the primary source of line photons, and $ {\hat R}(x,x',{\bf B})$
is a generic notation for a (6x6) angle-averaged frequency redistribution matrix,  its analytical expressions which vary according to different frequency domains
can be derived from the formulae given in Appendixes B and C of \citet{anusha2011c}. 

The reduced continuum source function may be cast into a form similar to Eq.(\ref{eq5}), replacing $ {\cal G}_l({\bf r}) $ by the primary source of 
photons in the continuum $ {\cal G}_c({\bf r}) $, and the frequency redistribution matrix $ {\hat R}(x,x',{\bf B})$ by $\delta(x-x') \hat{1} $ where $\hat{1} $ is the identity matrix.
  
  The reduced source function  ${\cal S}$ and intensity vector $ {\cal I}$ are complex quantities, in order to deal with real quantities
 \citet{frisch2007} introduced the following notations
 \begin{eqnarray}
 I_Q^{K,x}({\bf r}, \Omega, x)& = & Re( I_Q^{K}({\bf r}, \Omega, x)),\cr
 I_Q^{K,y}({\bf r}, \Omega, x)& = & Im( I_Q^{K}({\bf r}, \Omega, x)),
 \label{eq8}
 \end{eqnarray}
 and showed that ${\cal I}^r=(I_0^0,I_0^2,I_1^{2,x},I_1^{2,y},I_2^{2,x},I_2^{2,y})^T$  
 obeys a radiative transfer
 equation analogous to Eq. (\ref{eq2}) where the corresponding source term $ {\cal S}^r$ can be derived from expressions analogous to Eq. (\ref{eq5})
 with the
 matrix ${\hat \Psi}$ replaced by the matrix ${\hat \Psi}^r$, given in Appendix D of \citet{anusha2011c}.
 The derivation of the transfer equation for the real reduced  components of the radiation field relies on the symmetry property $( I_Q^{K})^*= (-1)^Q I_{-Q}^K$.

The reduced formalism for polarized transfer that we have summarized here  is well suited for efficient numerical solution, such as
Polarized Accelerated Lambda Iteration (PALI) or Jacobi methods, which may be applied in 3D media. Once the reduced components of the radiation field have
been obtained, one can easily compute the Stokes parameters of the radiation field using the relations given in the Appendix B of  \citet{frisch2007}.

\subsection{Boundary conditions for the reduced radiative transfer equation}

In the papers quoted above the radiative transfer problem was solved for self-emitting media in the absence of incident radiation. In \citet{milic2013}, 
the same formalism was used, but for computing  line-scattering polarization emerging from a circum-stellar disk illuminated by its central star. 
In order to self-consistently solve the reduced transfer equation one needs to cast the boundary conditions, i.e. the known incident radiation field, into the same
reduced basis. However, there is no unique way of deriving the reduced components $I_Q^K$ from a known radiation field described in terms of the Stokes parameters
(only the reversed operation is well defined). The reason is probably that the reduced formalism is well suited to describe the angular dependence of 
a radiation field arising from scattering processes, or from unpolarized and isotropic primary creation of photons, but it is not clear how to apply it to any arbitrary radiation. 
For problems where the medium is illuminated by an arbitrary radiation field, which can be polarized or unpolarized, we have to find a 
way of writing the boundary conditions for Eq. (\ref{eq2}). 

In the following we show that a consistent way of solving the problem, is to write the radiation field as the sum
of a transmitted radiation and a diffuse part. The diffuse radiation obeys a transfer equation like Eq. (\ref{eq2}), with an additional primary creation term arising
from the first scattering of the incident radiation in the medium. By the definition,  the diffuse radiation transfer problem has no incident radiation. The additional primary source term is easily cast into reduced components, because it originates from a scattering process. So the diffuse radiation may be computed in the reduced formalism presented
above and its Stokes parameters can then be derived. The full radiation is finally obtained in the Stokes formalism by adding the  Stokes components of the transmitted
radiation and the Stokes components of the diffuse radiation.

\section{Transmitted and Diffuse radiation fields}

We now deal with the polarized radiative transfer equation in the Stokes formalism, given by Eq. (\ref{eq1}), that we want to solve in the case where the medium is
illuminated by an  incident radiation field denoted by ${\bf I}_{inc}( {\bf r_0},\Omega, x)$, with known Stokes parameters $I_{inc}^i({\bf r}_0,\Omega, x)$, with $i=0,1,2,3$.
Like in \citet{chandra1960} and \citet{ivanov97} we shall write the solution of Eq. (\ref{eq1}) as the sum of two terms,
\begin{equation}
{\bf I}( {\bf r},\Omega, x)= {\bf I}_{inc}( {\bf r_0},\Omega, x)\exp (-\tau_{\Omega,x}({\bf r}))+ {\bf I}_d( {\bf r},\Omega, x),
\label{eq9}
\end{equation}
where $\tau_{\Omega,x}({\bf r})$ is the line optical depth from the boundary  at the location $ {\bf r}$ along the direction $\Omega$ at frequency $x$. The first term in the right-hand side
of Eq. (\ref{eq9}) is the transmitted radiation at depth ${\bf r}$, the second term is the so-called diffuse radiation field. Substituting Eq.(\ref{eq9}) in Eq. (\ref{eq1})
we easily derive the radiative transfer equation for ${\bf I}_d( {\bf r},\Omega, x)$, namely
\begin{equation} 
  {d {\bf I}_d( {\bf r},\Omega, x)\over ds}= - k_{tot}( {\bf r},x) [ {\bf I}_d( {\bf r},\Omega, x) - {\bf S}( {\bf r},\Omega, x)].
 \label{eq10}  
\end{equation}
The source term is the same as in Eq.(\ref {eq1}), it may be written in terms of the diffuse and transmitted radiation by 
substituting Eq. (\ref{eq9}) into the scattering integral.  The scattering integral is now the sum of two terms, both of which may be
expanded on the irreducible tensors ${\cal T}^ K_Q (i, \Omega)$. Thus, for the diffuse radiation field we can repeat the same
procedure  as described in the previous section, to cast the radiative transfer equation (\ref{eq10}) into a reduced form
similar to Eq.(\ref{eq2}), where the reduced source vector is given in Eq.(\ref{eq5}).

Here we write the line source vector ${\cal S}_l^r$, on its real form,  in terms of the diffuse field  ${\cal I}_d^r$, namely
  \begin{eqnarray}
 {\cal S}_l^r({\bf r},x) & = &
\int_{-\infty}^\infty  dx'  \int{d\Omega' \over 4\pi}{ {\hat R}(x,x', {\bf B})\over \phi(x)}
 {\hat \Psi}^r(\Omega'){\cal I}_d^r( {\bf r},\Omega', x') \cr
&+ & {\cal G}_l({\bf r}) + {\cal C}^r({\bf r},x) ,
  \label{eq11}
 \end{eqnarray}
 where $ {\cal C}^r=(C_0^0,C_0^2,C_1^{2,x},C_1^{2,y},C_2^{2,x},C_2^{2,y})^T$   is the 6-component additional source term, 
related to  the real and imaginary parts of  the complex quantities,
\begin{eqnarray} 
C_Q^K({\bf r},x)  & =  &{W_K\over  \phi(x)}\int_{-\infty}^\infty dx' \int{d\Omega' \over 4\pi} e^{ -\tau(\Omega',x')}\cr
 & &\sum_{j=0}^3R_{Q,j}^K(x,x',\Omega', {\bf B})I_{inc}^j({\bf r}_0,\Omega', x').
 \label{eq12}  
\end{eqnarray} 
The coefficients $W_K$ are constants which depend
on the quantum numbers of the line atomic levels. 
Analytical expressions of  $ {\cal C}^r$, as function of the Stokes parameters of the incident radiation are given in Appendix A.
The continuum source function in the reduced form has an expression similar to Eq. (\ref{eq11}), but for coherent Rayleigh
scattering.  



In non-magnetic media the redistribution matrix in Eq. (\ref{eq12}) reduces to 
\begin{equation}
R_{Q,j}^K(x,x',\Omega') = r^K(x,x') (-1)^{Q}{\cal T}^ K_{-Q}(j, \Omega'),
\label{eq21}
\end{equation}
with 
\begin{eqnarray}
r^K(x,x')={\Gamma_R\over \Gamma_R+\Gamma_I+\Gamma_E} 
& [& r_{II}(x,x') +\cr
 r_{III}(x,x') &. & {\Gamma_E-D^{(K)}\over \Gamma_R+\Gamma_I+D^{(K)}}].
\label{eq30}
\end{eqnarray}
We recall that in the presence of Hanle effect the redistribution matrix $ R_{Q,j}^K$ has different expressions according
 to different frequency domains \citep[see Appendix B in][]{anusha2011c}.
For the lines formed in dilute media (such as prominences) where collisions are
negligible, we have:

 In the frequency domain where $x(x+x') < 2 \nu_c^2(a)$ and $x'(x+x') < 2 \nu_c^2(a)$ 
\begin{eqnarray}
R_{Q,j}^K(x,x',\Omega', {\bf B}) &= &\sum_{Q'}{\cal M}_{Q,Q'}^K(B)
(-1)^{Q'}{\cal T}^ K_{-Q' }(j, \Omega')\cr
& .& r_{II}(x,x'),
\label{eq22}
\end{eqnarray}
and in the complementary domain,
\begin{equation}
R_{Q,j}^K(x,x',\Omega', {\bf B}) = 
(-1)^{Q}{\cal T}^ K_{-Q}(j, \Omega')
 r_{II}(x,x').
\label{eq23}
\end{equation}
The cutoff frequency $\nu_c$ between the line core and line wings depends on the value of the Voigt parameter $a$.  
It is usually on the order of a few Doppler widths.

The diffuse radiation field is zero at the boundaries of the medium. We can then solve its radiative transfer equation in reduced form
 and  derive its Stokes parameters  from the relations given in 
Appendix B of  \citet{frisch2007}. When the Stokes parameters of the diffuse field are computed  we obtain the Stokes parameters of the
total radiation field through Eq. (\ref{eq9}).

\section{Numerical tests}
The numerical solution of the reduced transfer problem for the diffuse radiation may be performed with the Jacobi iterative method 
introduced in \citet{anusha2011a}\citep[see also][]{milic2013}. 
The  additional source terms ${\cal C}^r({\bf r},x) $ and ${\cal C}_c^r({\bf r},x)$ have to be computed first on long characteristics from the boundaries to
each point of the spatial grid. We note that this significantly increases the computational work both because of using the long characteristics and the fact that angular integration must be performed on dense grids if high precision is to be achieved. For the numerical tests presented below we use a 100-point Gaussian quadrature for integration over $\theta$ while integration over azimuth is not needed because of the symmetry of the incident radiation. For this particular test case, using quadrature with 20 Gaussian points results in differences smaller than 1\%.

In all the tests we have used a slightly modified second order short characteristics \citep{OKSC} formal solution. A 2D-FBILI iterative procedure is used to obtain a self-consistent solution of the unpolarized NLTE problem \citep{MA13}. We then perform several $\Lambda$ iterations for the polarized part, with the same formal solver, in order to obtain the solution for the reduced source function.  

\subsection{Unpolarized incident radiation field}
In \citet{anusha2011a, anusha2011c} it is assumed that, if the incident radiation is  not polarized, one may
use as boundary conditions  for the reduced radiation field, the expression ${\cal I}_{\rm{inc}}= (I_{inc}, 0,0,0,0,0)^T$.
One can verify that in that case both this approach and the diffuse radiation field approach lead to identical expressions
for the Stokes parameters of the total radiation field. 

However the diffuse field approach allows us  to  take into account the anisotropy of the incident radiation in the computation of the polarized source function in a more accurate way, \emph{even in the case when the incident radiation is unpolarized}, because of the following: If one was to solve the radiative transfer problem one would have to specify $\cal{I}_{\rm{inc}}$ on a coarse angular grid. In the case of an anisotropic illumination, which also encounters abrupt changes (such as in the following examples), this would introduce errors in the computation of the reduced source vector. The accuracy is improved if one uses the averaging procedure described in \citet{Gouttebroze05} and subsequently in \citet{milic2013}. However, the averaging process is  constructed to accurately reproduce the $S_0^0$ component (identical to the scalar mean intensity  $J$), while the computation of the other components  is still not accurate. For externally illuminated objects of low and moderate optical thickness,
 the incident radiation makes for the dominant part of the reduced source function. This part is computed with a much higher precision with our approach because the averaging process is avoided, while the ``diffuse'' part of the computation can still be done with a coarser angular grid.

\subsection{Polarized  incident radiation field}
The formalism developed in  \citet{anusha2011a} does not allow to deal with the  boundary conditions in cases where the incident radiation is polarized. In that case we have to use the diffuse field approach. Here we consider \textbf{slabs which are effectively 1D (computations are performed with a 2D code)}, placed horizontally or vertically above the solar surface and illuminated by limb darkened, continuum radiation given by \citep[similar example as in:][]{landibook}:
\begin{equation}
 I_{\rm{inc}} = 1 - 0.95 (1 - \cos{\theta}) + 0.2 (1-\cos{\theta})^2.
\end{equation}
We consider a case where there is no continuum opacity, the line absorption profile is given by a Doppler function, and the line is formed by scattering processes only
and we assume that $W_0 = W_2=1$. The line-integrated optical depth in both cases is equal to unity. This creates some radiative transfer effects while still maintaining a 
large impact of the incident radiation. In both cases, the slab is located at height equal to $50 000$ km above the solar surface. Note that it introduces an additional geometrical 
anisotropy since the incident radiation is within a limited solid angle between the solar surface and the slab. As a coarse computational grid we use a grid with $12 \times 12$ angles where integrations over $\theta$ and $\varphi$ are done with Gaussian and trapezoid quadrature, respectively.

\begin{figure}
    \includegraphics[width=0.4\textwidth]{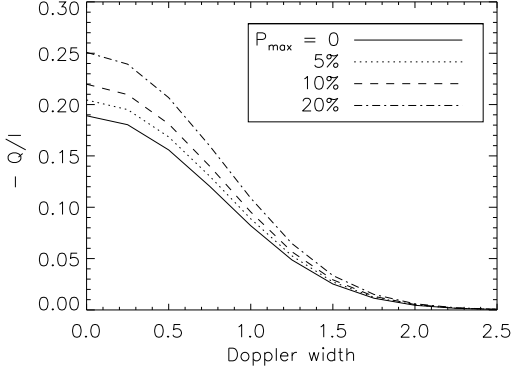}
    \caption{Q/I line profiles for different maximum degree of polarization of the incident radiation field emergent from a vertical slab}
    \label{line}
\end{figure}

\begin{figure}
    \includegraphics[width=0.4\textwidth]{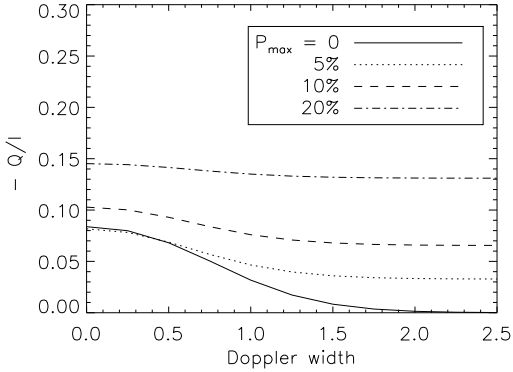}
    \caption{Same as Fig. \ref{line} except for a horizontal slab, $\mu = 0.6$}
    \label{line2}
\end{figure}

Fig. \ref{line} shows the emergent Q/I line profiles in the direction $\theta = \pi/2$, $\varphi = 0$, from a vertical slab (simple model of a solar prominence), illuminated by a
polarized radiation with the following angular dependence:
\begin{equation}
 Q = -P_{\rm{max}} \times I (1 - \cos^2\theta).
\end{equation}
We vary $P_{\rm{max}}$ between 0\% and 20\%. It is clearly seen that the degree of  polarization of the incident radiation significantly influences the polarization in the line. Similar effect can be seen in Fig. \ref{line2} in the case of the horizontal slab (simple model of a solar filament). Note that, in the latter case, the emergent radiation is a combination of the transmitted and scattered radiations.

\section{Conclusion}

We have demonstrated a consistent way of dealing with boundary conditions for polarized radiative transfer in the framework of the reduced intensity formalism. Our method is based on splitting the radiation field on a transmitted and a diffuse part. We then incorporate the incident radiation, given in the Stokes formalism, directly in the source function by performing high-precision angular integration and appropriate attenuation via long characteristics. This method is computationally slower but allows for higher precision, even in the unpolarized case and, more importantly, allows accounting for polarized incident radiation.

We have applied the method on simple, 1D slabs, which mimic prominences or filaments illuminated by the solar continuum radiation. We have shown that the degree of polarization of the incoming radiation may significantly influence the emergent line polarization from such structures. It is possible that the interpretation of UV lines formed by scattering of polarized continuum radiation in the upper layers of the Sun might require for such computational techniques. 

\acknowledgement{IM is thankful to Eiffel Scholarship of French Government and to COST MP1104 for supporting his stay in Nice. OA is thankful to the University of Nice-Sophia Antipolis for supporting her stay in Nice. This research is also partially funded by the Serbian Ministry of Science and Education through the project 176004, ``Stellar Physics''.}

\begin{appendix}
\section{Explicit expressions for the real reduced primary term due to the incident radiation}

We give here the analytical expressions for the additional source term in the RT equation for the diffuse radiation, in real form, $ {\cal C}^r=(C_0^0,C_0^2,C_1^{2,x},C_1^{2,y},C_2^{2,x},C_2^{2,x})^T$.

In the non-magnetic case 
\begin{eqnarray}
C_0^0 ({\bf r},x) & =  &{W_0\over  \phi(x)}\int_{-\infty}^\infty r(x,x')dx' \int{d\Omega' \over 4\pi}
I_{\rm{inc}}({\bf r}_0,\Omega', x') e^{ -\tau(\Omega',x')}\cr
C_0^2({\bf r},x) & =  &{W_2\over  \phi(x)}\int_{-\infty}^\infty r(x,x')dx' \int{d\Omega' \over 4\pi} e^{ -\tau(\Omega',x')}\cr
& & [{1\over 2\sqrt{2}}(3\cos^2\theta-1)I_{\rm{inc}}({\bf r}_0,\Omega', x')\cr 
& - &{3\over2\sqrt{2}}\sin^2\theta\, Q_{\rm{inc}}({\bf r}_0,\Omega', x') ]\cr
C_1^{2x}({\bf r},x) & =  &{W_2\over  \phi(x)}\int_{-\infty}^\infty r(x,x')dx' \int{d\Omega' \over 4\pi} e^{ -\tau(\Omega',x')}\cr
& & [-{\sqrt{3}\over 2}\sin\theta\cos\theta\cos\varphi\, I_{\rm{inc}}({\bf r}_0,\Omega', x') \cr
&- &{\sqrt{3}\over 2}\cos\theta\sin\theta\cos\varphi\,
Q_{\rm{inc}}({\bf r}_0,\Omega', x')\cr
& +& {\sqrt{3}\over 2}\sin\theta\sin\varphi\, U_{\rm{inc}}({\bf r}_0,\Omega', x')] ,\cr
C_1^{2y}({\bf r},x) & =  &{W_2\over  \phi(x)}\int_{-\infty}^\infty r(x,x')dx' \int{d\Omega' \over 4\pi} e^{ -\tau(\Omega',x')}\cr
& & [{\sqrt{3}\over 2}\sin\theta\cos\theta\sin\varphi\, I_{\rm{inc}}({\bf r}_0,\Omega', x')\cr
& + &{\sqrt{3}\over 2}\cos\theta\sin\theta\sin\varphi\,
Q_{\rm{inc}}({\bf r}_0,\Omega', x') \cr
&+& {\sqrt{3}\over 2}\sin\theta\cos\varphi \, U_{\rm{inc}}({\bf r}_0,\Omega', x')] ,\cr
C_2^{2x}({\bf r},x) & =  &{W_2\over  \phi(x)}\int_{-\infty}^\infty r(x,x')dx' \int{d\Omega' \over 4\pi} e^{ -\tau(\Omega',x')}\cr
& & [{\sqrt{3}\over 4}\sin^2\theta\cos2\varphi\, I_{\rm{inc}}({\bf r}_0,\Omega', x')\cr
& - &{\sqrt{3}\over 4}(1+\cos^2\theta)\cos2\varphi\,
Q_{\rm{inc}}({\bf r}_0,\Omega', x') \cr
&+& {\sqrt{3}\over 2}\cos\theta\sin2\varphi\, U_{\rm{inc}}({\bf r}_0,\Omega', x')] ,\cr
C_2^{2y}({\bf r},x) & =  &{W_2\over  \phi(x)}\int_{-\infty}^\infty r(x,x')dx' \int{d\Omega' \over 4\pi} e^{ -\tau(\Omega',x')}\cr
& & [-{\sqrt{3}\over 4}\sin^2\theta\sin 2\varphi\, I_{\rm{inc}}({\bf r}_0,\Omega', x')\cr
& +&{\sqrt{3}\over 4}(1+\cos^2\theta)\sin 2\varphi\,
Q_{\rm{inc}}({\bf r}_0,\Omega', x') \cr
&+& {\sqrt{3}\over 2}\cos\theta\cos 2\varphi\, U_{\rm{inc}}({\bf r}_0,\Omega', x')],
\label{eqA1}
\end{eqnarray}
\end{appendix}

In the presence of a magnetic field, for lines formed in dilute media, the same magnetic matrix as given in \citet{frisch2007} can be applied to ${\cal C}^r$ in order to account for the Hanle effect.

\bibliography{faurob}
\bibliographystyle{aa}

\end{document}